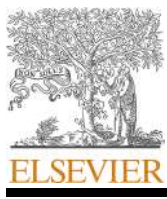
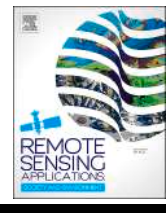

# Monitoring of tectonic deformation in the seismic gap of the Mentawai Islands using ALOS-1 and ALOS-2

Pakhrur Razi [a,b,*], J.T.S. Sumantyo [c], Ming Yam Chua [d], Ganefri [e], Daniele Perissin [f], Takeo Tadono [g]

[a] *Physics Department, Math, and Natural Science Faculty, Universitas Negeri Padang, Indonesia*
[b] *Center of Disaster Monitoring and Earth Observation, Universitas Negeri Padang, Indonesia*
[c] *Center of Environmental Remote Sensing, Chiba University, Japan*
[d] *School of Electrical Engineering and Artificial Intelligence, Xiamen University Malaysia, Malaysia*
[e] *Electrical Department, Faculty of Engineering, Universitas Negeri Padang, Indonesia*
[f] *Lyle School of Civil Engineering, Purdue University, West Lafayette, USA*
[g] *Earth Observation Research Center Space Technology Directorate I, Japan Aerospace Exploration Agency (JAXA), Japan*

ABSTRACT

Ever since the last occurrence of a significant earthquake in the Mentawai megathrust zone in 2000, no significant earthquake events have been recorded, which, according to the earthquake repetition cycle, suggests that the zone is a potential epicenter of future earthquakes. The southern and northern parts of the zone have been struck by a significant earthquake with magnitude M > 8.0; however, in the potential location of the Mentawai Islands, earthquake energy has not been released. This research shows the tectonic activity, velocity, and shift that occurred owing to the thrust of the plate. The information is a vital reference for estimating the epicenter of the earthquake whose energy has not yet been released. We analyzed the tectonic characteristics according to the synthetic aperture radar data and geodetic global positioning system observations. The results show that the Pagai Islands are experiencing consistent tectonic deformations. The northern region of North Pagai and the Northern region of South Pagai are experiencing significant subsidence, while the southwest (SW) region of North Pagai and the south segment of South Pagai are experiencing significant uplift. The government and local authorities can use this information as a guide for developing strategies for disaster preparation.

## 1. Introduction

The Sumatra megathrust zone is one of the active seismic areas that are consistently threatened by earthquake hazards. The zone is located in the western part of Sumatra Island, Indonesia. According to the history of earthquake occurrence along the Sumatra seismic belt in the past few decades, the worst earthquake happened in 2004, with magnitude M 9.15; followed by an M 8.6 earthquake in 2005 in the northern region; an M 8.4 earthquake in 2007; and an M 7.9 earthquake in 2000 in the southern region of the Sumatra megathrust zone. Ever since the last occurrence of a significant earthquake in 2000, no significant earthquake incidents have been recorded. The Mentawai Islands are a group of three small islands: Siberut Island, Sipura Island, and the Pagai (North and South) Islands. The earthquake with the highest recorded magnitude of M 7.9 occurred at Eastern Pagai Island in September 2007. An M 7.2 earthquake occurred at North Pagai in February 2008, an M 7.0 earthquake occurred at Sipora Island in September 2007, and an M 7.8 earthquake occurred at the southwest of South Pagai in October 2010.

The Pagai Islands can be divided into North Pagai Island and South Pagai Island. From May 2007 to December 2019, 97 earthquakes (with magnitude ≥ 5) occurred on the Pagai Islands. The occurrence frequency dropped yearly: from 33 in 2007 to 20 in 2008 and 2–7 each year from 2009 to 2018, with all classified as shallow-depth earthquakes. However, in 2019, the number of shallow-depth earthquakes with magnitudes M > 5 increased to 11 in the northern region of South Pagai Island. This could be a sign of an impending huge earthquake because the area belonged to the ''supercycle'' of the last 200 years (Philibosian et al., 2017). Most of the previous studies on earthquakes on the Mentawai Islands were conducted using only global positioning system (GPS) data (Sieh and Natawidjaja, 2000), (Han et al., 2014), (Wang et al., 2018) and seismic and bathymetry data (Singh et al., 2010). However, in this research, we combined synthetic aperture radar (SAR) and GPS data for analysis.

There is a need to monitor and map the tectonic deformation on the Pagai Islands, which could help in estimating the possible earthquake epicenter and hence predicting the possible threat of a tsunami. Early damage prevention work could be conducted by the local authorities to minimize the losses caused by earthquakes and tsunamis. The state-of-the-art SAR technology and GPS can be used to detect surface changes on the earth with a detection accuracy of up to sub-millimeter resolution. In this research, we applied differential interferometric SAR (D-InSAR) and persistent scatterer interferometry (PSI) techniques for monitoring tectonic deformation on the Pagai Islands (see Fig. 6), using the SAR data provided by Japan Aerospace Exploration Agency from Advanced Land Observing Satellite Phased Array-type L-band Synthetic Aperture (ALOS-1 PALSAR) and ALOS-2 PALSAR. The data were recorded from September 26, 2009, to December 30, 2010, and February 24, 2015, to September 3, 2019, respectively. Both techniques can detect changes in the earth's surface with centimeter- and millimeter-level accuracy, respectively. Both techniques combined with remote sensing applications have been widely applied in landslide detection (Razi et al., 2018), subsidence detection (Aobpaet et al., 2013), volcano eruption detection (Hooper et al., 2004), building stability detection (Silvia LiberataUllo et al., 2019), (Zhu et al., 2020), earthquake modeling (Jonsson, 2002), and fault detection. Moreover, the Sumatran GPS array (SuGAr) network data were used to validate the results from the SAR satellite processing and data analysis.

## 2. Materials and methods

### 2.1. Study area and satellite dataset

The study area lies in North Pagai Island and South Pagai Island, which are parts of the Mentawai Islands, with coordinates 100° 12′ E and 2° 48' S. The study area is in the megathrust zone, where the Indo-Australia plate is moving beneath the Eurasian plate. The Pagai Islands and some small surrounding islands were formed owing to the subduction mechanism of the Indo-Australia to Eurasia plate (Philibosian et al., 2017), (Scholz, 2019). According to digital elevation model (DEM) data, the topography of North Pagai Island and South Pagai Island can be classified into coastal land, low land, middle land, and upland. Most of the area in the Pagai Islands is flat on the coast and elevates to the mainland, with the highest topography of 300 m above sea level. The coastal land area starts from the coastal line and rises into a 0°%–3°% slope zone toward the mainland. The coastal areas are lowlands, marshes, and muddy regions. The low-land areas have choppy topography with slopes of 3%–8% and are free from tidal influences. The middle land borders the low land and the hills with a slope zone of 5%–8%. The residents use these areas for agricultural purposes. The topography and geology formation of the Pagai Islands are shown in Fig. 1.

According to the geological structure, the soil in the Pagai Islands is dominated by latosol, podsolic, and alluvial soil. Latosol covered ∼70% of North Pagai Island, and podzolic covered 15% of the island. The rest areas are covered by alluvial deposits, which are distributed along the coastal line. According to the geomorphology, the area along the coastal line consists of terraced sandy coral reefs and shell fragments with a slope of 2°–7°.

According to the earthquake data from January 2007 to August 2020, the epicenters of the earthquakes on the Pagai Islands were distributed in several areas (Fig. 3). However, one of the earthquakes was not recorded as an earthquake event with magnitude M 7.0 or higher (red square area). Most earthquake events in 2019 belong to the red square area in Fig. 3, with the highest magnitude being M 6.0, which occurred on February 2, 2019. The intensity of earthquakes around the area suggests that a large earthquake epicenter will occur along the fault line of Pagai Island.

Fig. 3 shows the concentrations and distributions of the epicenters of four significant earthquakes on the Pagai Islands. The first significant earthquake, with magnitude M 7.9, occurred on September 12, 2007, at 2.625°S and 100.841°E in the 35 km depth. Another earthquake, with magnitude M 7.0, occurred on the island the next day (September 13, 2007) at 2.130°S and 99.627°E, at a depth of 22 km. On February 25, 2008, another earthquake with magnitude M 7.2 struck the island, at 2.486°S and 99.972°E, at a depth of 22 km. Then, on October 25, 2010, another earthquake with magnitude M 7.8 occurred on the Pagai Islands, at 3.487°S 100.082°E, at the 20.1 km depth and triggered a tsunami.

According to the bathymetry map in Fig. 4, North Pagai and South Pagai and their inter-island strait feature deep-sea waters, with depths in the range of 0–750 m and 0–200 m in the western and eastern parts of the island, respectively. The sea depth deepens to the western subduction zone, which reaches 6000 m, while in the east, a basin forms between the Mentawai Islands and Sumatra, with a sea depth of ∼1300 m. Moreover, the study area is surrounded by two tectonic faults. First, a subduction zone (megathrust) may be found in the Pagai Island's western region, which faces the Indian Ocean. The fault is a result of the thrust of the Indo-Australia plate beneath the Eurasia plate, resulting in a depth of up to 70 km. In 2010, a significant earthquake with a magnitude of 7.8 occurred in the area and was followed by a huge tsunami. Between the Mentawai Islands and Sumatra Island lies the Mentawai fault. The fault resulted from the oblique subduction of Indo-Australia to Eurasia that spreads from the South to the North of the Mentawai Islands.

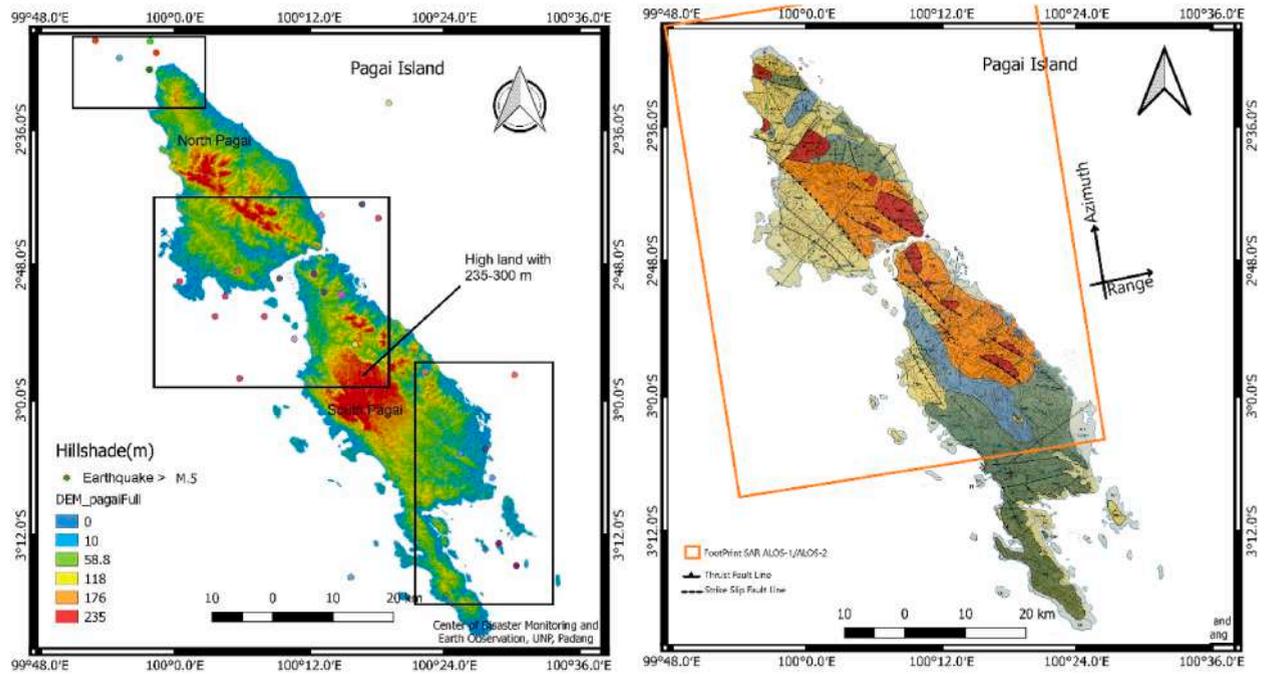

**Fig. 1.** Topography and geology formations of North Pagai and South Pagai, part of the Mentawai Islands, West Sumatra, Indonesia. Left: the RGB color indicates the high level of the area above sea level, and squares black are concentrations of earthquake epicenters; right: the color indicates soil type, and square orange is the footprint of ALOS-1/ALOS-2 SAR data. (For interpretation of the references to color in this figure legend, the reader is referred to the Web version of this article.)

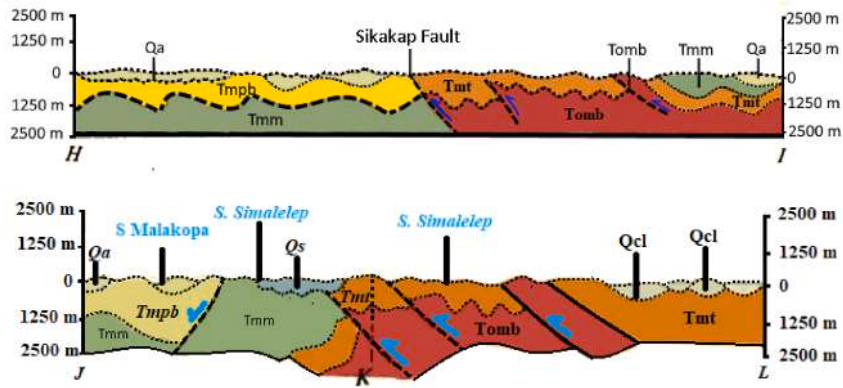

**Fig. 2.** Subsurface formation of the Pagai Islands and fault location. (Above): subsurface formation of North Pagai Island, (below): subsurface formation of South Pagai Island. The blue arrow is the fault direction, and the dotted line is the fault line. (For interpretation of the references to color in this figure legend, the reader is referred to the Web version of this article.)

Fig. 4 presents the epicenter of earthquakes along the Sumatra megathrust zone. The highest magnitude of earthquakes is concentrated in the southern and northern regions of the Sumatra megathrust zone. However, in recent years, no earthquake has been recorded for the area between the regions (orange square area). This area is called a seismic gap and is located on the Mentawai Islands. The last huge earthquakes in this area occurred on February 10, 1797, with magnitudes M 8.4–8.9, and on November 25, 1833, with magnitudes M 8.8–9.2, which triggered a tsunami. In recent years, earthquakes with magnitudes M 7.6 and 7.8 occurred near the Mentawai seismic gap, on September 30, 2009, and October 25, 2010, respectively. Both earthquakes have not fully released the accumulated energy since their last release in 1797 and 1833.

The SAR satellite datasets used in this study are tabulated in Tables 1 and 2. The data had a high resolution of 3 × 10 m in range and azimuth. Two satellite missions were considered, and the satellites for both missions featured vertical transmit and vertical receive (HH) polarizations, with a microwave center frequency of 1.27 GHz (23.6 cm wavelength). The data acquisition was dated from June 26, 2009, to September 03, 2019.

The ALOS-1 and ALOS-2 satellite datasets featured the same wavelength, frequency, orbit direction, beam mode, and polarization. For the analysis of the ALOS-1 dataset, the scene acquired on December 27, 2009, was selected as the master scene, and the others were used as slave scenes. With the chosen master and slave scenes, the minimum and maximum normal baselines ($B_n$) were 35 and 1043 m, respectively, and the maximum temporal baseline was 369 days. The scene captured on July 25, 2017, was selected as the master for the ALOS-2 dataset, with a maximum temporal baseline of 881 days and the longest normal baseline of 477 m. A small nor-mal baseline will produce more accurate results, whereas a small temporal baseline will enhance the coherence (Razi et al., 2018) of the interferogram.

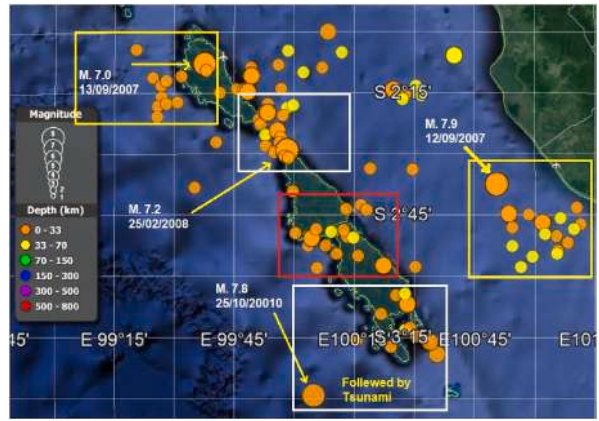

**Fig. 3.** The epicenter of earthquakes with magnitudes of >5 in the Pagai Islands from May 2007 to August 2020 (USGS).

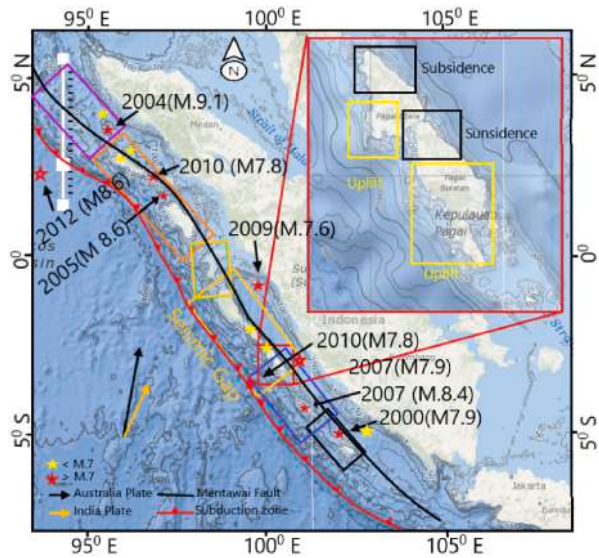

**Fig. 4.** Bathymetry map of Sumatra Island: The red square is the bathymetry map of the study area on the Pagai Islands. Inset: The yellow square is the uplifted area, and the black squares are the subsidence based on SAR observation. (For interpretation of the references to color in this figure legend, the reader is referred to the Web version of this article.)

**Table 1**
Satellite dataset of ALOS-1.

| Satellite mission | Acquisition time | $\lambda$ (cm) | $f_0$ (GHz) | orbit | Beam mode | Pol. | Bn (m) | Bt (days) |
| --- | --- | --- | --- | --- | --- | --- | --- | --- |
| ALOS-1 | 26/06/2009 | 23.6 | 1.27 | Ascending | SM3 | HH | −250 | 184 |
| | 26/09/2009 | 23.6 | 1.27 | Ascending | SM3 | HH | 35 | 92 |
| | 27/12/2009 | 23.6 | 1.27 | Ascending | SM3 | HH | 0 | 0 |
| | 11/02/2010 | 23.6 | 1.27 | Ascending | SM3 | HH | −98 | 45 |
| | 29/06/2010 | 23.6 | 1.27 | Ascending | SM3 | HH | 186 | 183 |
| | 14/08/2010 | 23.6 | 1.27 | Ascending | SM3 | HH | 26.7 | 229 |
| | 29/09/2010 | 23.6 | 1.27 | Ascending | SM3 | HH | 74.7 | 275 |
| | 14/11/2010 | 23.6 | 1.27 | Ascending | SM3 | HH | 1043 | 321 |
| | 20/12/2010 | 23.6 | 1.27 | Ascending | SM3 | HH | 495 | 369 |

The SuGAr dataset (Table 3) used in this research was observed from January 2, 2019, to December 31, 2019.

### 2.2. Methods

#### 2.2.1. D-InSAR technique

D-InSAR is a powerful technique for detecting changes in the earth's surface. This technique has been widely used not only in land deformation applications but also for earthquake observation (Razi et al., 2020), (Fang et al., 2019), volcano eruption monitoring (Huang et al., 2017), landslide mapping, and land subsidence mapping (Simons and Rosen, 2015), (Gao et al., 2019). The D-InSAR principle works by generating an interferometric phase (interferogram) through the complex multiplication of master and slave SAR images with the same orbit but acquired at different times (see Fig. 5). The interferometric phase can provide details on alterations to the earth's surface over a wide area if the earth's surface movement (with reference to the master scene) can be measured up to sub-millimeter-level accuracy.

**Table 2**
Satellite dataset of ALOS-2.

| Satellite mission | Acquisition time | λ (cm) | $f_0$ (GHz) | orbit | Beam mode | Pol. | Bn (m) | Bt (days) |
|---|---|---|---|---|---|---|---|---|
| ALOS-2 | 22/02/2015 | 23.6 | 1.27 | Ascending | SM3 | HH | 200 | 881 |
|  | 23/02/2016 | 23.6 | 1.27 | Ascending | SM3 | HH | 477 | 517 |
|  | 04/10/2016 | 23.6 | 1.27 | Ascending | SM3 | HH | −29 | 293 |
|  | 25/07/2017 | 23.6 | 1.27 | Ascending | SM3 | HH | 0 | 0 |
|  | 06/03/2018 | 23.6 | 1.27 | Ascending | SM3 | HH | 445 | 224 |
|  | 29/05/2018 | 23.6 | 1.27 | Ascending | SM3 | HH | 299 | 308 |
|  | 04/09/2018 | 23.6 | 1.27 | Ascending | SM3 | HH | 65 | 406 |
|  | 22/01/2019 | 23.6 | 1.27 | Ascending | SM3 | HH | 290 |  |
|  | 03/09/2019 | 23.6 | 1.27 | Ascending | SM3 | HH | −55 | 770 |

**Table 3**
SuGar GPS data.

| Station | Type | Frequency | Latitude | Longitude | Location |
|---|---|---|---|---|---|
| SMGY | Geodetic | Dual frequency | −2.614490 | 100.103000 | Saumanganya |
| SLBU | Geodetic | Dual frequency | −2.766340 | 100.000000 | Silabu |
| BSAT | Geodetic | Dual frequency | −3.076710 | 100.285000 | Bulasat |
| PRKB | Geodetic | Dual frequency | −2.966600 | 100.400000 | Parak Batu |

The interferometric phase difference between the two acquisition times can be expressed as (Razi et al., 2020)

$$\varphi_1 = \frac{4\pi R}{\lambda} \text{ and } \varphi_2 = \frac{4\pi (R + \Delta R)}{\lambda} \quad (1)$$

$$\Delta \varphi = \varphi_1 - \varphi_2 = \frac{4\pi \Delta R}{\lambda} \quad (2)$$

where $S_1$ and $S_2$ are satellite positions at different acquisition times, $R$ is the slant range of the earth's surface to the satellite, and $\lambda$ is radar wavelength. These two components contribute to the interferometric phase in D-InSAR processing.

$$\Delta \varphi_{1,2}^{DinSAR} = \Delta \varphi_{1,2} - \Delta \varphi_{1,2}^{flat} - \Delta \varphi_{1,2}^{height} \quad (3)$$

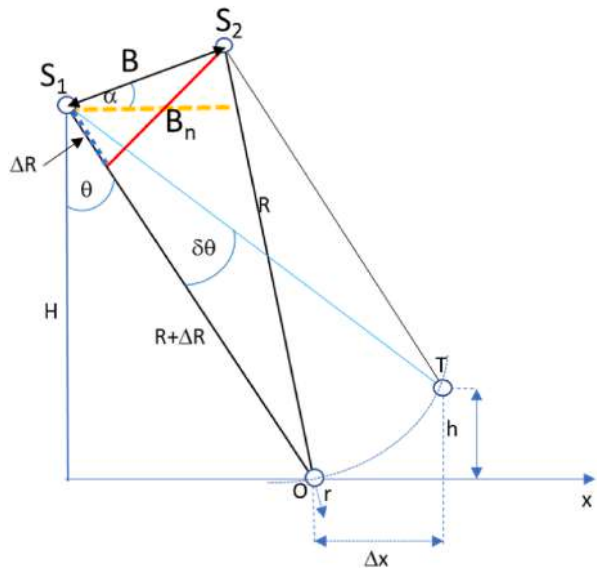

**Fig. 5.** SAR geometry.

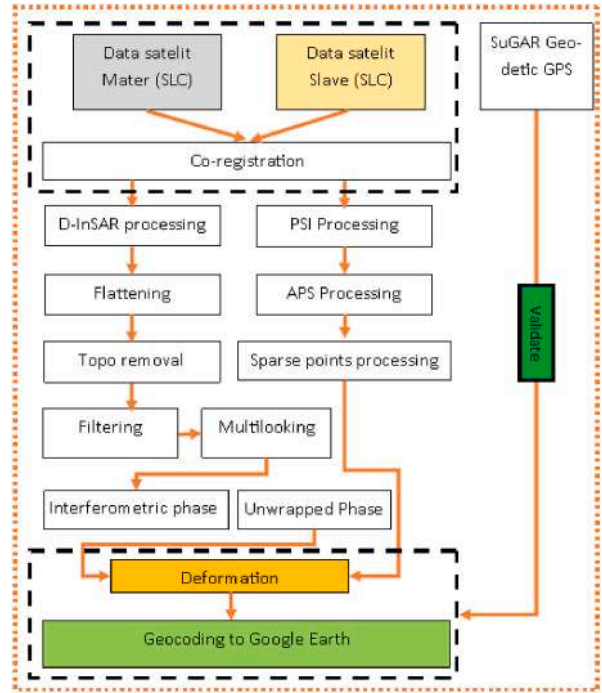

**Fig. 6.** Schematic of D-InSAR and PSI processing for the tectonic deformation model.

The atmospheric phase and noise phase effects can be removed through filtering and multi-look processing techniques. The flat and height phase components can be estimated from the DEM and then removed. After the interferometric phase is obtained, an unwrapped technique is applied to obtain the real value of deformation.

*2.2.2. PSI–SAR technique*

PSI is an advanced D-InSAR technique that can minimize geometric and temporal decorrelation and atmospheric disturbance. The PSI technique exploits multiple images over the same area and orbit with respect to a single master image (Razi et al., 2019). The master image is selected by maximizing the stack coherence value, $\gamma^m$, of the combination of the images.

$$\gamma^m = \frac{1}{N} \sum_{i=1}^{N} g\left(B_n^{m,s}, B_{n\ max}\right) \cdot g\left(B_t^{m,s}, B_{t\ max}\right) \cdot g\left(B_{DC}^{m,s}, B_{DC\ max}\right) \tag{4}$$

where $-N$ is the total number of slave images; $-B_n^{m,s}$ is a normal baseline between the master and slave (m); $-B_{n\ max}$ is the normal baseline maximum (m) calculated according to the effect of the decorrelation, and it is dependent on the range, azimuth bandwidth, and radar wavelength; $-B_t^{m,s}$ is the temporal baseline (days); $-B_{t\ max}$ is the temporal baseline maximum (days); $-B_{DC}^{m,s}$ is the Doppler centroid baseline (the mean Doppler centroid frequency difference).

The components that contribute to the formation of the interferometric phase in PSI processing can be expressed as (Kampes, 2006), (Ferretti, 2014), and (Perissin, 2016).

$$\Delta\varphi_{m,s}(T) = \Delta\varphi_{m,s}^{flat}(T) + \Delta\varphi_{m,s}^{height}(T) + \Delta\varphi_{m,s}^{disp}(T) + \Delta\varphi_{m,s}^{atm}(T) + \Delta\varphi_{m,s}^{nois}(T) \tag{5}$$

The master image is indexed by *m*, and the slave image is indexed by *s*. The term $\Delta\varphi_{m,s}^{flat}(T)$ is the flat terrain phase component related to the earth's curvature. It can be estimated from the orbital data of the satellite during scene acquisition.

$$\Delta\varphi_{m,s}^{flat}(T) = \frac{4\pi}{\lambda} \frac{B_n}{R_m \tan\theta} \tag{6}$$

$\Delta\varphi_{m,s}^{height}(T)$ is the topography height component related to the inaccuracy of the DEM; it is used as a reference, which corresponds to the height of the target and normal baseline. It can be formulated as

$$\Delta\varphi_{m,s}^{height}(T) = \frac{4\pi}{\lambda} \frac{B_n}{R_m} \frac{\Delta h(T)}{\sin\theta} \tag{7}$$

where $-B_n$ is the perpendicular distance between master $S_m$ and slave $S_s$ images, and it can also be called a normal baseline; $-R_m$ is the distance in the slant range from the satellite to reference point O; $-\Delta h(T)$ is the elevation of the point target $T$ relative to a reference point O on the ground; $-\theta$ is the angle between the nadir to the beam of the antenna in the range direction (off-nadir angle). $\Delta\varphi_{m,s}^{disp}(T)$ is the relative displacement of target $T$ to a reference point on the ground and temporal baseline $B_t$. In a linear model, the displacement of the target point can be expressed as

$$\Delta\varphi_{m,s}^{disp}(T) = \frac{4\pi}{\lambda} B_t \Delta v(T) \tag{8}$$

where $-\Delta v(T)$ is the velocity of the target T; $-\Delta\varphi_{m,s}^{atm}(T)$ is an atmospheric phase component, which can be calculated and estimated using a residual model; $-\Delta\varphi_{m,s}^{nois}(T)$ is the noise contribution from SAR data conversion and thermal noise; it is estimated and removed from the residual model.

The first, fourth, and fifth interferometric phase components can be estimated and removed, and then, the remaining component can be expressed as

$$\Delta\varphi_{m,s}(T) = \Delta\varphi_{m,s}^{height}(T) + \Delta\varphi_{m,s}^{disp}(T) = \Delta\varphi_{m,s}(T) = \Delta\varphi_{m,s}^{height}(T) + \Delta\varphi_{m,s}^{disp}(T) \tag{9}$$

where $\lambda$ is the wavelength of a signal transmitted by the radar system.

### 2.2.3. Geodetic GPS SuGar network observation

The SuGar network is a series of GPS stations located along the west coast of Sumatra and its islands in the Indo-Australian subduction zone to Eurasia. This network measures the deformation due to tectonic movements and earthquakes and then sends the data to the central stations located at LIPI Bandung and EOS Singapore. To obtain the magnitude of the velocity and the direction and magnitude of the shift, the processing is conducted using the GAMIT/GLBOK software. Both GAMIT and GLBOK can be used to estimate coordinates and shifts of GPS stations and obtain physical quantities. The amount of deformation at each point of the monitoring station is calculated in the coordinate system using the equation:

$$dE_{12} = (E_2 - E_1) \; x \; 111320 \; metre \tag{10}$$

$$dN_{12} = (N_2 - N_1) \; x \; 111320 \; metre \tag{11}$$

where - $dE_{12}$ is an eastward movement of SuGar station; - $dN_{12}$ is a northward movement of SuGar station; - $E_1$ is the initial GPS in the longitudinal position; - $E_2$ is the final GPS in the longitudinal position; - $N_1$ is the initial latitude of a GPS location; - $N_2$ is the final latitude of a GPS location.

Through the calculation of the resultant vector of the movement of the GPS station, the direction and magnitude of the shift of the GPS station can be obtained.

$$R = \sqrt{(dE_{12})^2 + (dN_{12})^2 + 2dE_{12}dN_{12}\cos\theta} \tag{12}$$

The direction of the resultant displacement can be obtained by calculating the magnitude of the angle

$$\tan\alpha = \frac{dR}{dN_{12}} \tag{13}$$

where $R$ is the resultant of station GPS shifted, $\theta$ is an angle formed between $dE_{12}$ and $dN_{12}$.

The velocity of movement at each station can be calculated using the equation:

$$v_h = \sqrt{v_e^2 + v_n^2} \tag{12}$$

where - $v_h$ is the horizontal velocity, - $v_e$ is the velocity in the east direction - $v_n$ is the velocity in the north direction.

## 3. Results

### 3.1. Differential interferometric SAR observation

Through the D-InSAR technique, the tectonic deformation and its value were obtained via complex multiplication between two SAR images (one master scene and one slave scene) acquired at different times. The process was performed for every pair of SAR images in the datasets from ALOS-1 and ALOS-2. The interferometric phase for each pair of SAR images was acquired from 2009 to 2019. During the processing, the phase noise that influenced the phase of the radar signal was removed via multi-look processing (3 × 3 in the range and the azimuth direction) and boxcar filtering.

To obtain the deformation map and the real displacement value, the phase was unwrapped. The deformation and displacement map values for each pair of SAR images are shown in Fig. 7.

From July 26, 2009, to September 26, 2009 (Fig. 7a), the northern regions of North Pagai Island and South Pagai Island were uplifted by approximately −100 mm (close to the satellite). Moreover, the southwest region of North Pagai Island subsided by ~100 mm (far away from the satellite). In addition, some small islands in the northwest of South Pagai Island were uplifted by ∼200 mm and subsided by up to 300 mm in a local area in the northern part of South Pagai. A similar deformation was observed on February 11, 2010, according to pair scenes from December 27, 2009, and February 11, 2010 (Fig. 7b). However, the deformation value increased by ∼160 mm in the south region of North Pagai Island and ∼80 mm in South Pagai Island. In the northern part of North Pagai Island, the area subsided by 160 mm. These significant deformations were observed from September 26, 2009, to July 29, 2010 (Fig. 7c). The area on North Pagai Island was uplifted by 500 mm, and several areas on the island subsided by about 400–800 mm. In another section, most of South Pagai Island subsided by up to 400–800 mm, although the west part of South Pagai Island was uplifted by ∼400 mm. An earthquake that occurred on September 30, 2009, with an epicenter at 0°43′N, 99°58′E and magnitude M 7.6 in the 81 km of depth, was the cause of the deformation during this time. The earthquake destroyed numerous buildings and killed people in Padang City and the Padang Pariaman district.

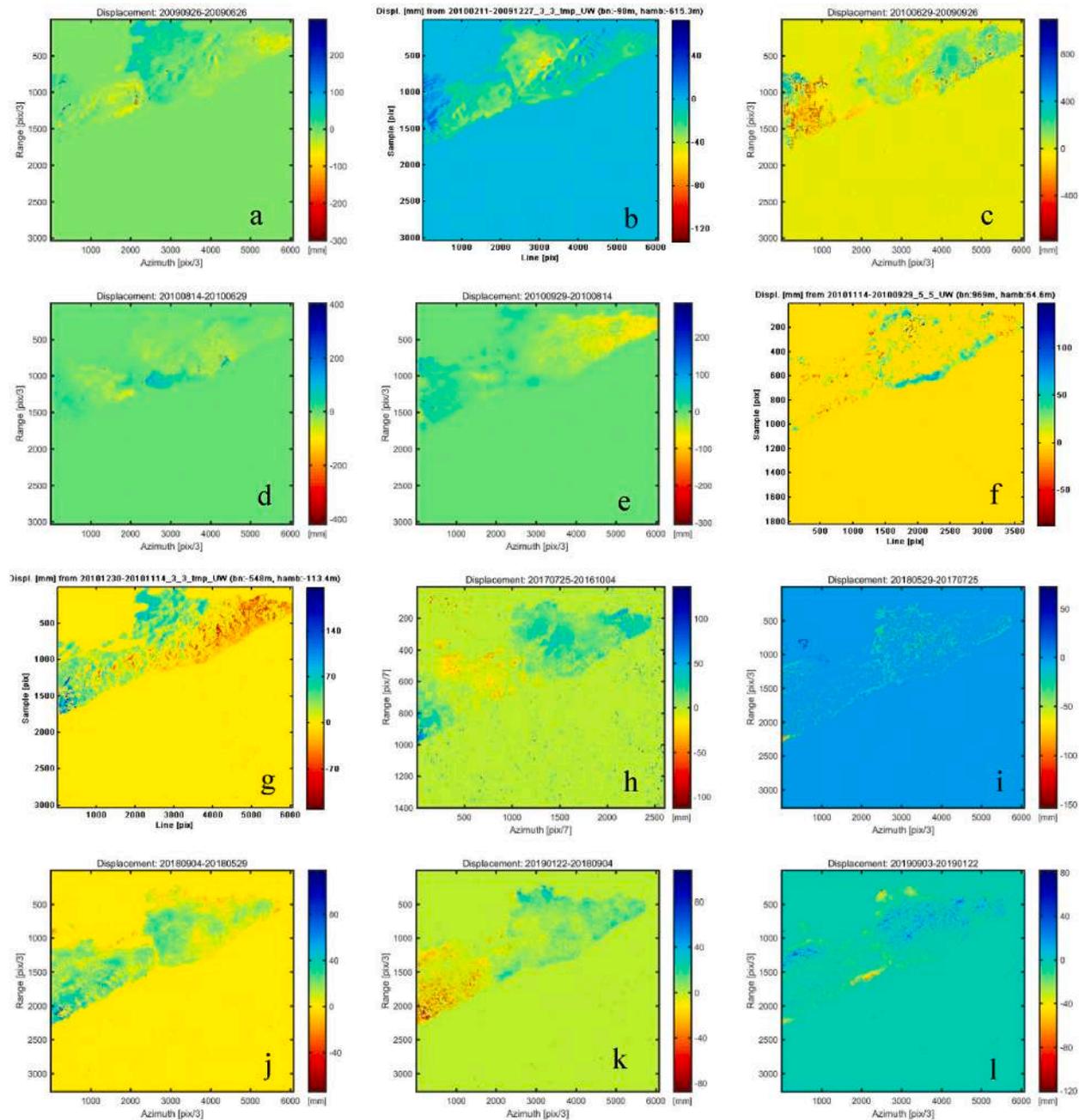

**Fig. 7.** Deformation map and displacement value of the Pagai Islands from 06 to 26–2009 to 09-03-2019.

For the observations from July 29, 2010, to August 14, 2010 (Fig. 7d), the coastal line areas in the northwest and southeast of North Pagai Island were uplifted by ∼200 mm. The rest of the areas on the island subsided up to 100 mm. However, the south region of South Pagai Island was uplifted by ~100 mm. A similar trend was observed from August 14, 2010, to September 29, 2010 (Fig. 7e). Most of the areas on North Pagai Island subsided by ~100 mm, except in the southwest and south regions. The coastal line of that island was uplifted to a height of 110 mm. However, from September 29, 2010, to November 14, 2010 (Fig. 7f), the mainland of the Pagai Islands could not be observed owing to a lack of radar signals. The information could only be extracted for areas along the coastal line on South Pagai Island and North Pagai Island. At this time, most of the coastal line area in western Pagai Island was uplifted by ~150 mm, but the eastern region experienced a subsidence of ~100 mm, although some small areas in the location were uplifted. This condition was caused by an earthquake on October 25, 2010, with magnitudes M 7.6, M 6.3, and M 5.0, which caused a tsunami. Similarly, the scenes from November 14, 2010, to December 30, 2010 (Fig. 7g), showed that most of the areas in North Pagai Island experienced subsidence, while most of the areas in South Pagai Island were uplifted.

The next series of observations were based on the ALOS-2 dataset, which began on February 22, 2015, with the same reference point as the ALOS-1 dataset. In the pair scene from October 04, 2016, to July 25, 2017 (Fig. 7h), North Pagai was uplifted by 50–100 mm. The northern region of South Pagai subsided by 25–100 mm, and the southern part of the island was uplifted by ~100 mm. In the scene from July 25, 2017, to May 29, 2018 (Fig. 7i), some locations in the western region of the Pagai Islands, including the small islands, were uplifted by ~50 mm. Nevertheless, the mainland subsided by ~100 mm. One months after that, on June 25, 2018, an earthquake with magnitude M 5.0 occurred on the mainland of the island, with the epicenter at 2.813°S, 100.207°E in the 29.15 km depth. The earthquake affected the tectonic deformation in the area along the western coastal line of the Pagai Islands, which subsided by ~80 mm. However, the eastern region of the island was uplifted by ~100 mm. The same tectonic deformation was observed using the dataset from June 29, 2018, to September 4, 2018 (Fig. 7j).

The scene from September 04, 2018, to January 22, 2019, showed another significant deformation in South Pagai Island, where the entire area subsided by ~80 mm. However, North Pagai Island was uplifted by ~100 mm. Ten days after last observation, on February 2, 2019, a series of earthquakes struck the island, with the epicenter located around South Pagai Island. Nineteen earthquakes were recorded, six with magnitudes M 5.2, M 6.0, M 5.3, M 5.9, M 5.8, and M 5.1 (see red square in Fig. 2), and 13 with magnitudes M 4.1 to M 4.9. The earthquakes caused subsidence in most of the areas of South Pagai Island. The magnitude along the coastal line on western and eastern Pagai Island reached 120 mm. Some uplifting also occurred in the middle region of North Pagai Island. Fig. 7k shows the map of the deformation caused by a series of earthquakes. From January 22, 2019, to September 3, 2019, most of the coastal line in the western part of the Pagai Islands subsided by up to 120 mm, with the middle region of the land being uplifted by a maximum magnitude of 80 mm (Fig. 7l).

According to this study, it can be deduced that the Indo-Australian plate activity in the west of the Pagai Islands moves under the Eurasian plate and then thrusts some parts of the island, which causes elevation and subsidence. The plate shatters into multiple segments after the movement reaches its critical point, causing earthquakes.

### 3.2. PSI–SAR observation

The PSI–SAR technique was also applied to obtain the tectonic deformation trend for the whole observation time. The technique can provide additional information on land movement velocity, trend, and displacement. To reduce the atmospheric and noise effects on SAR data, atmospheric phase screen processing was applied. To obtain the persistent scatterer (PS) candidate, an amplitude stability index of 0.75–1 was chosen.

Fig. 8 shows the PS point distribution that indicates the tectonic deformation trend of the Pagai Islands. The velocity rate of the PS point on the line of sight in the area ranges from −300 to +300 mm/year. A PS with a negative velocity indicates that the surface of the area moves away from the satellite (subsidence), while a PS with a positive velocity indicates that the land surface moves toward the satellite (elevation). The color gradients from blue to dark red and blue to dark blue indicate increasing deformation rates in subsidence and uplift, respectively.

According to observations from June 26, 2019, to December 20, 2010, subsidence occurred in most regions of the Pagai Islands, reaching 300 mm/year, but the north area of South Pagai and a small segment of North Pagai experienced uplift. Furthermore, from February 22, 2015, to September 03, 2019, the subsidence areas in the Pagai Islands were (i) the northern region of South Pagai Island; (ii) the southern region of North Pagai Island; and (iii) the northern region of North Pagai Island, with minimum and maximum rates of −30 to −200 mm/year. The southern region of South Pagai Island and the southwest region of North Pagai Island, including the surrounding small islands, experienced uplifting. The minimum and maximum rates of change were from +20 to +100 mm/year. Nevertheless, some small areas on the mainland underwent subsidence in the western and southeastern regions of South Pagai Island.

### 3.3. Geodetic GPS SuGar network observation

There are four GPS stations on the Pagai Islands: SMGY, SLBU, BSAT, and PRKB. The GPS data observations were used for additional analysis in determining the velocity, magnitude, and direction of land movement on the Pagai Islands. Moreover, they were used for comparison and correction of observations based on radar SAR. The trend of the movement of the Pagai Islands is depicted in Fig. 9.

From June 26, 2009, to October 24, 2010, both the SMGY and SLBU stations moved to the northeast, and then reversed significantly to southwest on October 25, 2010, owing to the M 7.8 earthquake in the western part of South Pagai. The earthquake caused the area at the station to subside by ~50 mm. However, at the BSAT and PRKB stations, the movement toward the northeast was not too large, or it could have approached saturation before turning around. However, after the earthquake, the area at stations SMGY and SLBU experienced greater subsidence of 150 and 30 mm, respectively, because station BSAT was close to the earthquake epicenter.

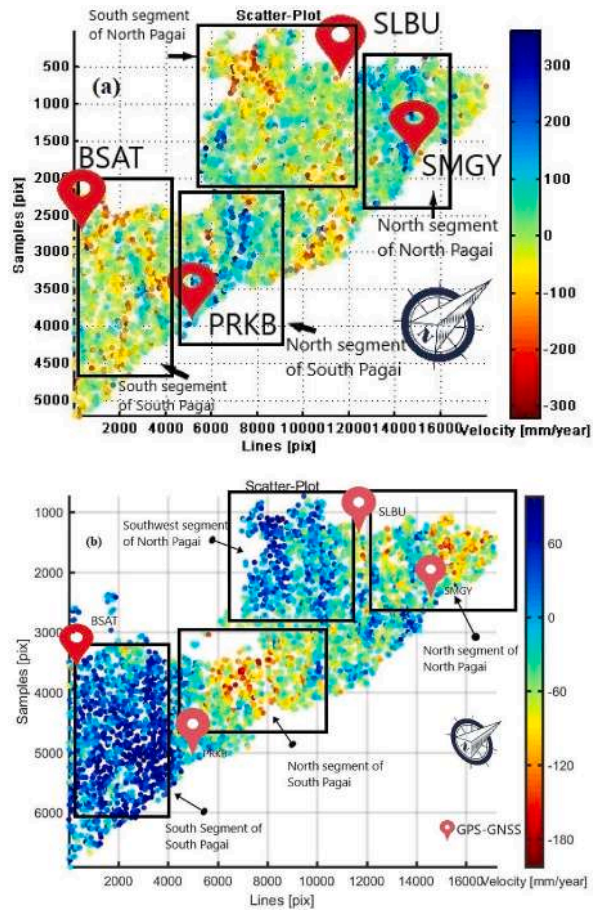

**Fig. 8.** The tectonic deformation trend of the Mentawai Islands is represented by the PS distribution. (a) from 06 to 26–2009 to 12-20-2010; (b) from 02 to 22–2015 to 09-03-2019.

Owing to the limitations of GPS data, the next detailed observation was conducted from 01 to 02–2019 to 12-31-2019. The movement and deformation trends that occurred in the four stations are depicted in Fig. 10.

Fig. 10 depicts the land movement of the Pagai Islands in a time series from DoY (day of year) 002 to 365 in 2019 for both stations SMGY and SLBU. Most of the day featured an uptrend that is indicated by a movement in the northern direction. However, on DoY 031, a co-seismic jump occurred in both stations owing to the earthquake on February 2, 2019. This led to an area on the western side of North Pagai with a subsidence of up to 40 mm and an uplift of up to 30 mm on the eastern side of North Pagai. For this event, both the BSAT and PRKB stations were inactive. After DoY 200, the fourth station featured an uptrend in the north and east axes, and subsidence for the up axis (yellow graph). The detailed velocity and movement for the fourth station are listed in Table 4.

## 4. Discussion

Generally, the GPS-measured plate movement velocity of Indo-Australia to Eurasia at the western region of the Mentawai Islands is 43–60 mm/year (Qin and Singh, 2018). We extracted, analyzed, and monitored the tectonic deformation caused by plate movement using radar technology based on ALOS-1 PALSAR 1, ALOS-2 PALSAR 2, and geodetic GPS. The observation period was from June 2009 to October 2019. According to the analysis, from February 2015 to September 2019, the plate movement of Indo-Australia beneath Eurasia in the Mentawai Islands led to subsidence in the northern region of North Pagai Island (Pasapuat, Simagandjo, Silaoinan, and Tapuraukat) and the northern region of South Pagai Island (Seai), with a velocity of up to 200 mm/year. Conversely, the southwestern region of North Pagai Island (Betumonga, Silabu) and the southwestern and southern regions of South Pagai Island (Makalo, and Malakopa) were uplifted. The measured rates of both land deformations (i.e., subsidence and uplift) were higher than the velocity rate of plate movement that occurred on the western side of the Mentawai Islands. Furthermore, from June 2009 to December 2010, the northern (Saumanganyak), southwestern (Batumonga), and eastern parts (Sikakap) of North Pagai Island and the western part (Malakopak) of South Pagai exhibited subsidence of up to 300 mm. After the earthquake on October 25, 2010, with magnitude M 7.8, the northern part of North Pagai and the western and eastern parts of South Pagai Island subsided by up to 100 mm and were uplifted to 90 mm in the western region of North Pagai Island.

The results of processing using ALOS-1 PALSAR-1 (Fig. 8a) were compared with that of processing using the SuGAr network installed in North Pagai (SLBU and SMGY) and South Pagai (BSAT and PRKB). The SLBU GPS station in North Pagai recorded a subsi-dence trend until 2010, and there were no data from February to May 2010. From mid-2010, the subsidence trend continued until October 2010 and suddenly ceased owing to the occurrence of an M 7.8 earthquake. A similar trend was also detected by the SMGY and BSAT GPS stations. However, the PRKB GPS station located in eastern South Pagai detected an uplift. The fourth geodetic GPS deformation trend was depicted in Fig. 9.

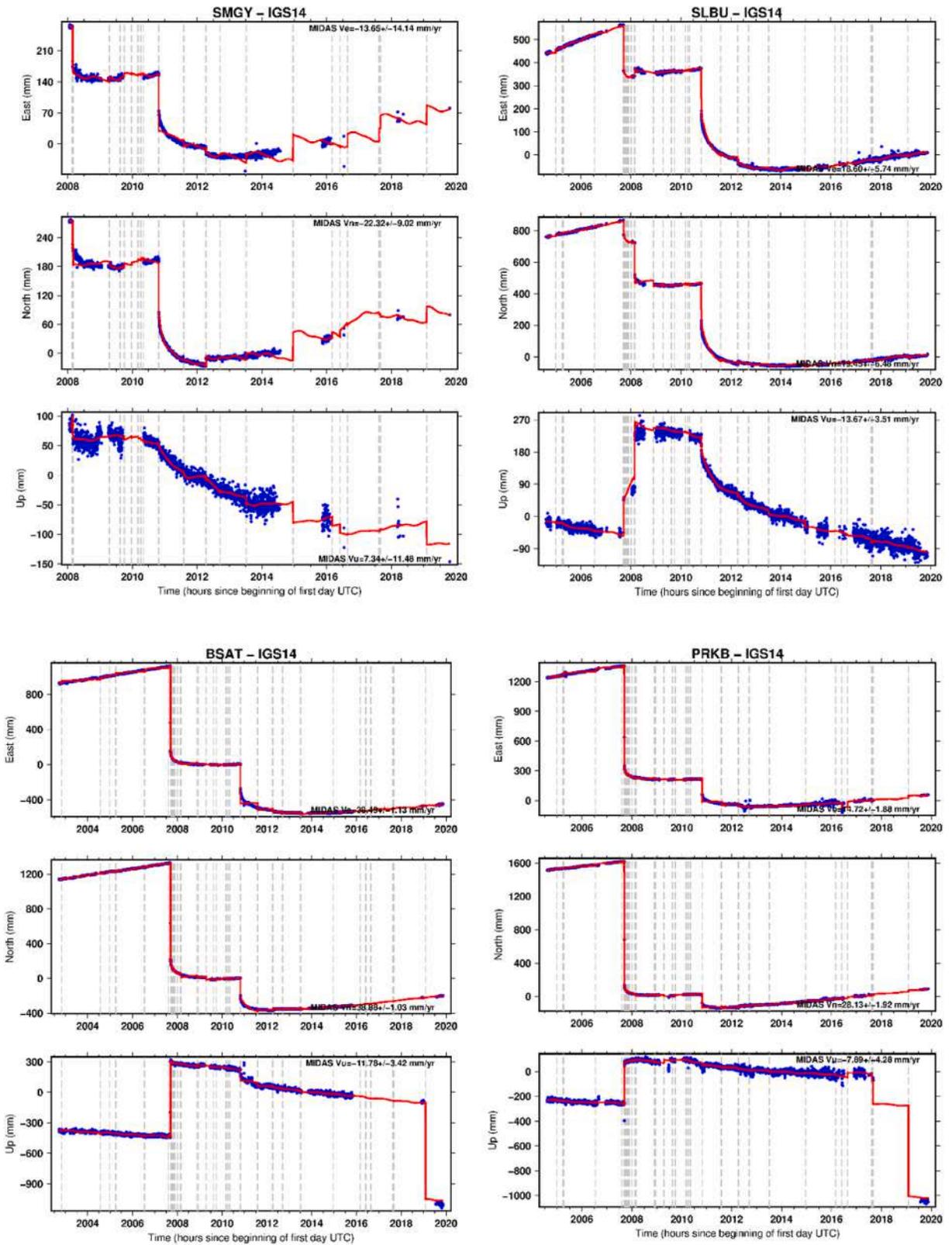

**Fig. 9.** Deformation trends in the southwest (SLBU) and north (SMGY) of North Pagai and southwest (BSAT) and east (PRKB) of South Pagai Island from 2002 to 2019, observed using geodetic GPS of the SuGAr network.

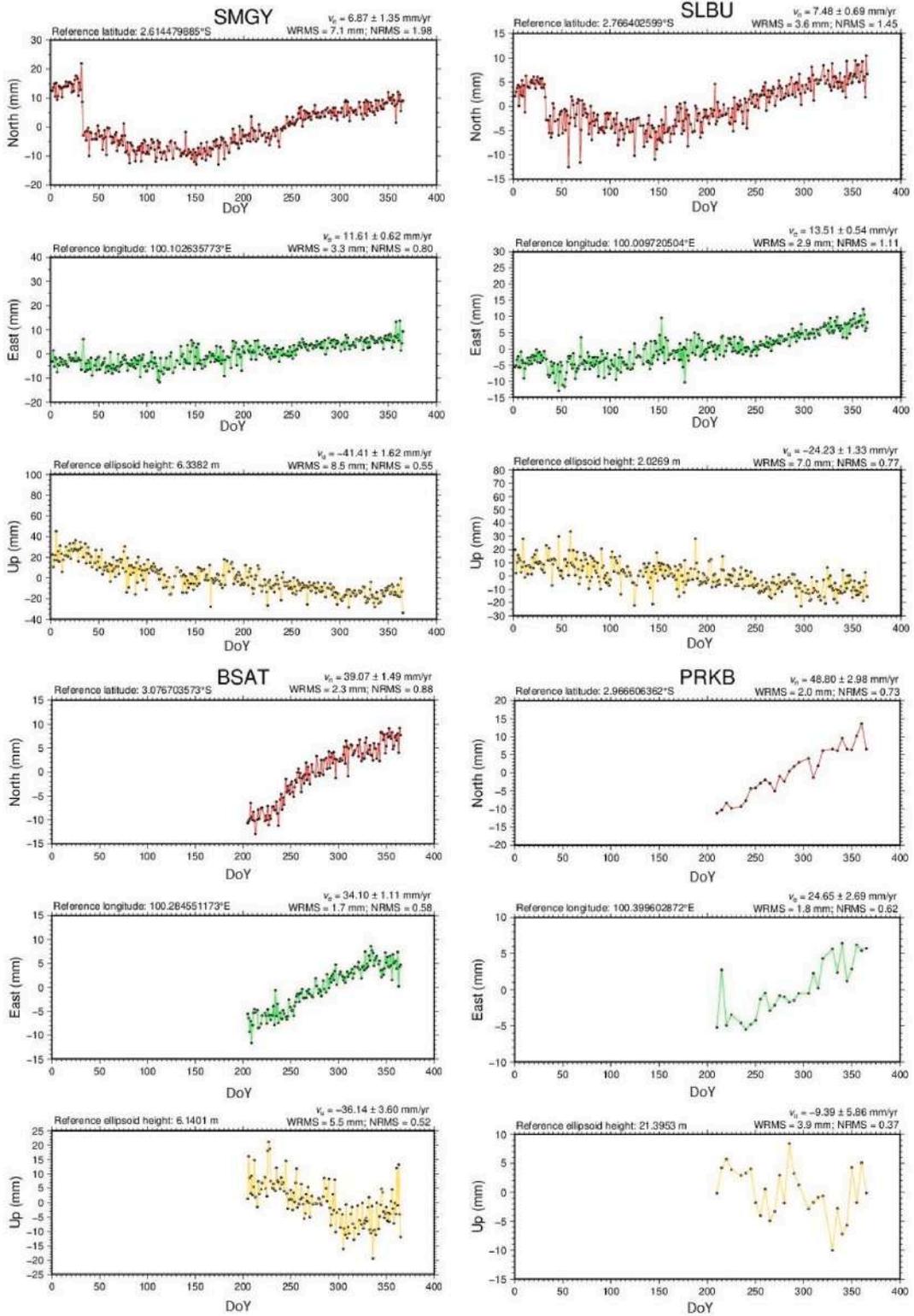

**Fig. 10.** The land movement of the Pagai Islands in a time series was based on four geodetic GPS stations.

**Table 4**
The velocity rate of land movement of the Pagai Islands from 01 to 02–2019 to 12-31-2019 per day.

| GPS array station | Land velocity (mm/yr) | | Horizontal velocity (mm/yr) | Vertical velocity (mm/yr) | α |
|---|---|---|---|---|---|
| | Vn | Ve | vh | Vu | |
| **SMGY** | 6.87 | 11.61 | 13.49 | −41.41 | S 74.09° E |
| **SLBU** | 7.48 | 13.51 | 15.44 | −24.23 | N 71.62° E |
| **BSAT** | 39.07 | 34.10 | 51.86 | −36.14 | N 29.17° E |
| **PRKB** | 48.80 | 24.65 | 54.67 | −9.39 | N 31.79° E |

The observations from February 22, 2015, to September 23, 2019, were analyzed using the dataset from ALOS-2 PALSAR-2. The study area exhibited a deformation trend (Fig. 8b) in which most of the areas continually underwent subsidence. However, some areas in the southwest of North Pagai and the southern region of South Pagai were uplifted. This was confirmed by the GPS data installed on the island (Fig. 9). However, in 2019, most areas in the coastal line were subsiding (SAR data [Fig. 7], and GPS data [Fig. 10], Fig. 11, and Table 4). This shows the existence of plate movement activity from the Indo-Australian to the Eurasian plate in the western part of the Mentawai Islands (Lange et al., 2018). In addition, the BSAT station (southwest of South Pagai) recorded higher movement in the horizontal and vertical directions than the other stations, attributable to the occurrence of several earthquake epicenters in this area. The PRKB station is more likely to move toward the northeast, although it also experiences slight subsidence. Regardless, both the SMGY and SLBU stations are more likely to experience subsidence than move toward the northeast. According to recent earthquake records, the PRKP location has fewer earthquake epicenters compared with the BSAT and SLBU locations. Therefore, the PRKP location has the potential to be the epicenter of an earthquake in the future.

Considering the earthquake repetition-based cycle, we have now entered the 200-year earthquake cycle, even though in the 2000s, several earthquakes with magnitude M > 5 on the Richter scale have been concentrated in the three zones of the Mentawai Islands (Fig. 3). However, the total energy released is not comparable to the accumulated energy released in the last 200-year cycle in 1797 and 1833, which led to M 8.4–8.6 and M 8.8–9.2 Richter-scale earthquakes, respectively. This means that accumulated strain energy is stored around the area; thus, early mitigation and preparation work can be started in the area to face potential future disasters.

## 5. Conclusion

In this work, SAR data from ALOS-1 and ALOS-2 were processed using both D-InSAR and PSI techniques and geodetic GPS data of the SuGAR network. The data were used to monitor the tectonic deformation in the seismic gap of the Mentawai Islands (Pagai). Every pair of results from SAR data processing and geodetic GPS observation from June 2009 to September 2019 showed significant tectonic deformation on the Pagai Islands. According to the results of both techniques, the northern Pagai region is more likely to experience land subsidence than it is to move toward the northeast as it usually does. The area along the coastline experienced a signifi-cant decline, particularly on the eastern, northern, and western sides of North Pagai. The same phenomenon also occurred in South Pagai, particularly in the north and west. A large amount of land subsidence and a horizontal shift toward the northeast occurred in the western region of South Pagai at the same time, while the eastern region of South Pagai exhibited a noticeable movement toward the northeast but little subsidence. These occurrences are due to the effects of the Simalelep fault, a back thrust fault, and the Malakopa fault, a normal fault; both faults divide the western and eastern sides of South Pagai (Fig. 2).

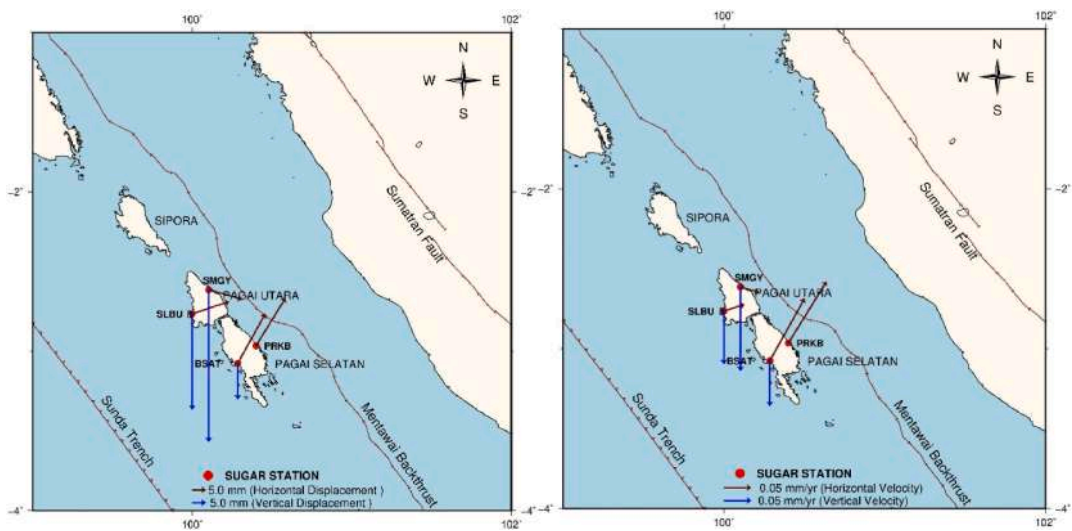

**Fig. 11.** (Left): Map of the displacement vector, and (Right): Map of the velocity vector of the Pagai Islands in 2019.

One of the reasons North Pagai is depressed and experiences land subsidence is the eight times greater migration to the north of South Pagai than to the north of North Pagai. The locking on the northern side is likely the cause of the sluggish speed northward from North Pagai (Sipora and Siberut). According to these findings, the region that may provide a location for the release of energy-accumulating stresses in the seismic gap begins from North Pagai to the north. However, the combined monitoring outcomes based on radar SAR and geodetic GPS satellite data are restricted to observed changes on the surface. Considering observational data from the subsurface region can lead to more comprehensive outcomes in the future.

According to the historical data, several large earthquakes occurred in the subduction zone in the Mentawai Islands in 1350/1388, 1658/1703, and 1797/1833, which are estimated to have recurred every 200 years, whereas today is almost the peak of the recurrence period. In this study, during our 10-year observation, we found that the western part of South Pagai Island up to North Pagai Island subsided and that some areas in the southwest of South Pagai and North Pagai uplifted (SAR: Fig. 8a and b; GPS: Figs. 10 and 11, Table 4). The observed phenomena serve as an alert to the Indonesian government, local authorities, and the public to get ready to face a potential earthquake on the west coast of West Sumatra.

**Data availability statement (DAS)**

AW data from ALOS-1 and ALOS-2 cannot be shared because of the term of the collaboration agreement with the Japan Aerospace Exploration Agency (JAXA), Japan. Optical satellite data from LANSAT-7 and Sentinel-2 can be accessed using https://earthexplorer.usgs.gov/ and GPS data can be accessed at http://sugar.geotek.lipi.go.id.

**Author contributions**

Conceptualization, P.R, MYC, and JTS.S; methodology, P.R, D.P; software, P.R. and D.P.; validation, P.R. and JTS.S.; formal analysis, JTS.S.; investigation, P.R.; data curation, P.R. and JTS.S.; writing—original draft preparation, P.R, and MYC.; visualization, P.R, G.; supervision, JTS.S.; project administration, AO.

**Acknowledgments**

The author would like to thank the Japan Aerospace Exploration Agency (JAXA), which supported the SAR data in the 2nd Earth Observation Research Collaborative Agreement between the Japan Aerospace Exploration Agency (JAXA) and the Center of Disaster Monitoring and Earth Observation, Universitas Negeri Padang, with contract number 19/JAXA/S1MD No. 0509001 and PI number ER2A2N011. The Ministry of Education, Culture, Research, and Technology, Indonesia, with grant number 409/UN35.13/LT/2021. Also, to the USGS, Google Earth, BNPB, and BPBD of West Sumatra province for external data support.

**Funding**

This research received external funding from the Japan Aerospace Exploration Agency, Japan, under grant number 0509001. The Ministry of Education, Culture, Research, and Technology, Indonesia, with grant number 409/UN35.13/LT/2021, and Universitas Negeri Padang with grant number 1419/UN35.13/LT/2020.

**Ethical statement for remote sensing applications: Society and environment**

Hereby, I am Pakhrur razi consciously assure that for the manuscript Monitoring of Tectonic Deformation in the Seismic Gap of the Mentawai Islands using ALOS-1 and ALOS-2 the following is fulfilled:

1) This material is the authors' own original work, which has not been previously published elsewhere.
2) The paper is not currently being considered for publication elsewhere.
3) The paper reflects the authors' own research and analysis in a truthful and complete manner.
4) The paper properly credits the meaningful contributions of co-authors and co-researchers.
5) The results are appropriately placed in the context of prior and existing research.
6) All sources used are properly disclosed (correct citation). Literally copying of text must be indicated as such by using quotation marks and giving proper reference.
7) All authors have been personally and actively involved in substantial work leading to the paper, and will take public responsibility for its content.

The violation of the Ethical Statement rules may result in severe consequences.

I agree with the above statements and declare that this submission follows the policies of Remote Sensing Applications: Society and Environment as outlined in the Guide for Authors and in the Ethical Statement.

**Declaration of competing interest**

The authors declare that they have no known competing financial interests or personal relationships that could have appeared to influence the work reported in this paper.

**Data availability**

Data will be made available on request.